\documentclass[12pt]{article}
\usepackage{graphicx} 

\def\bbR{\mathbb{R}}
\pdfoutput=1
\usepackage{graphicx}
\usepackage{amsmath, amssymb, amsfonts, amsthm, mathrsfs, setspace, fullpage}
\usepackage{enumitem}
\usepackage{bm,booktabs,float,url}
\usepackage{xcolor}
\usepackage{algorithm}
\usepackage{stmaryrd}
\usepackage{algorithmic}
\usepackage[square, authoryear]{natbib}
\usepackage{cases}
\usepackage[colorlinks,
            linkcolor=black,
            anchorcolor=blue,
            citecolor=blue]{hyperref}
\newtheorem{theorem}{Theorem}

\newtheorem{lemma}{Lemma}

\newtheorem*{lemma-non}{Lemma}
\newtheorem{asup}{Assumption}

\def\ba{\mathbf{a}}
\def\be{\mathbf{e}}
\def\cE{\mathcal{E}}
\def\bbE{\mathbb{E}}
\def\cG{\mathcal{G}}
\def\cH{\mathcal{H}}
\def\bI{\mathbf{I}}
\def\bF{\mathbf{F}}
\def\bh{\mathbf{h}}

\def\bS{\mathbf{S}}
\def\bP{\mathbf{P}}

\def\bbP{\mathbb{P}}
\def\bbR{\mathbb{R}}

\def\bs{\mathbf{s}}
\def\bp{\mathbf{p}}
\def\cV{\mathcal{V}}

\def\bx{\mathbf{x}}

\def\bX{\mathbf{X}}
\def\bZ{\mathbf{Z}}
\def\bA{\mathbf{A}}

\def\bB{\mathbf{B}}
\def\bD{\mathbf{D}}
\def\bQ{\mathbf{Q}}

\def\cX{\mathcal{X}}

\def\bfbeta{\bm{\beta}}
\def\bSigma{\bm{\Sigma}}
\def\bmu{\bm{\mu}}

\def \TP{\operatorname{TP}}
\def \FP{\operatorname{FP}}
\def \TN{\operatorname{TN}}
\def \FN{\operatorname{FN}}
\def\bOmega{\mathbf{\Omega}}

\def\tr{\operatorname{tr}}
\def\argmin{\operatornamewithlimits{argmin}}

\newcommand{\ind}{\perp\mkern-9.5mu\perp}

\allowdisplaybreaks

\newcommand{\blind}{1}

\begin{document}

\if1\blind
{
  \title{Nonparametric learning of heterogeneous graphical model on network-linked data}
  \author{
    Yuwen Wang$^\dag$, Changyu Liu$^\dag$, Xin He$^\ddag$, and Junhui Wang$^\dag$ \\
    [10pt]
    $^\dag$Department of Statistics \\
    Chinese University of Hong Kong \\
    $^\ddag$School of Statistics and Management \\
    Shanghai University of Finance and Economics
   }
  \date{} 
  \maketitle
} \fi

\if0\blind
{
  \bigskip
  \bigskip
  \bigskip
  \begin{center}  
    {\LARGE  Nonparametric learning of heterogeneous graphical model on network-linked data}
  \end{center}
  \medskip
} \fi

\onehalfspacing
\begin{abstract}

Graphical models have been popularly used for capturing 
conditional independence structure in multivariate data, which are often built upon independent and identically distributed observations, limiting their applicability to complex datasets such as network-linked data. This paper proposes a nonparametric graphical model that addresses these limitations by accommodating heterogeneous graph structures without imposing any specific distributional assumptions. The proposed estimation method effectively integrates network embedding with nonparametric graphical model estimation. It further transforms the graph learning task into solving a finite-dimensional linear equation system by leveraging the properties of vector-valued reproducing kernel Hilbert space. Moreover, theoretical guarantees are established for the proposed method in terms of the estimation consistency and exact recovery of the heterogeneous graph structures. Its effectiveness is also demonstrated through a variety of simulated examples and a real application to the statistician coauthorship dataset. 
\end{abstract}

{\em Keywords:} Conditional independence, Gaussian graphical model, heterogeneous data, network model, score matching 

\newpage    
\doublespacing

\section{Introduction}

Graphical models provide a powerful framework for modeling conditional independence structures in multivariate data and have attracted increasing attention in both academia and industry \citep{Lauritzen1996, Meinshausen2006, Yuan2007, friedman2008sparse, liu2009nonparanormal, li2020high}. Although success has been widely reported, most existing graphical models heavily rely on the assumption of independent and identically distributed observations, and thus are not applicable to analyze datasets with heterogeneous distribution or dependent samples, such as network-linked data.

Network-linked data has been commonly encountered in modern real-world applications, where the network connects observations that may exhibit dependence structure and heterogeneous distributions. To address the non-tensor structure in such datasets, various statistical models for network-linked data have been developed under specific scenarios, such as classification \citep{tang2013universally} or regression \citep{li2019prediction, le2022linear, zhang2022joint}. In particular, \citet{li2020high} introduced a network-linked Gaussian graphical model that accommodates heterogeneous mean vectors while assuming a common graph structure across all nodes. Another related line of research is the time-varying Gaussian graphical model \citep{zhou2010time,Kolar2011,lu2018post}, where the timeline can be regarded as a simple chain network and the graph structures are allowed to evolve over each time node. It is important to note that most of the aforementioned graphical models are built upon the Gaussian assumption, and the graph structure is then fully captured by its precision matrix. Although simple in estimation, verifying the Gaussian assumption can be challenging in many medium-scale real applications.

This paper proposes an efficient method to learn a nonparametric graphical model that accommodates heterogeneous graph structures over each node without imposing any restrictive distributional assumptions. The proposed method estimates the node-wise heterogeneous graph models as functions of the embedding vectors of the newtork in a nonparametric fashion, where the nonparametric graphical model is motivated by the equivalence between conditional independence and the zero pattern of the second-order derivative of the log density. We then estimate the derivative function via score matching \citep{hyvarinen2005estimation,zhou2020nonparametric} in a vector-valued reproducing kernel Hilbert space (RKHS; \cite{kimeldorf1971some}) induced by a new decomposable kernel function. By leveraging the representer theorem and the derivative reproducing property, the proposed method leads to a closed-form solution, where the representer coefficients are derived by solving a linear equation system, enabling efficient estimation of large-scale heterogeneous graphical models. 

We also conduct rigorous theoretical analysis that establishes the consistency of the proposed graphical model in exactly recovering the heterogeneous graph structures without imposing any explicit distributional assumptions. The analysis involves a careful study of the vector-valued RKHS and requires some novel theoretical treatments on the Hilbert-Schmidt operators. We also provide some theoretical examples, to showcase that all required assumptions are verified and the non-asymptotic convergence rate of the proposed method can be established.  The effectiveness of the proposed method is further supported by numerical comparison against some existing competitors in several simulated examples as well as a real application to the statistician coauthorship dataset.

The rest of the paper is organized as follows. In Section \ref{sec:mod}, we present the nonparametric heterogeneous graphical model on network-linked data. Section \ref{sec_estimation} introduces an efficient two-step learning algorithm. Theoretical guarantees, including estimation consistency and exact graph recovery for the proposed graphical model, are presented in Section \ref{sec_theo}. Section \ref{sec:num} presents simulation studies and a real application to the statistician coauthorship dataset. Section \ref{sec:dis} gives some final concluding remarks. Proofs of the main theoretical results are included in the Appendix, and other additional technical details, including auxiliary Lemmas S1–S11, are available in the online Supplementary Material.

\paragraph{Notations.} We denote $[k]=\{1, \dots, k\}$ for any positive integer $k$. For a vector $\ba$, the
$p$-norm is denoted as $\|\ba\|_p$. For a matrix $\mathbf{M}=(m_{ij})$, the max norm is $\|\mathbf{M}\|_{\max}=\sup_{i, j} |m_{ij}|$, the Frobenius norm is $\|\mathbf{M}\|_F = \big ( \sum_{i,j} m_{ij}^2 \big )^{1/2}$, the spectral norm is $\|\mathbf{M}\|_2=\sup_{\|\ba\|_2=1}\| \mathbf{M}\ba\|_2$, and the $(2, \infty)$-norm is $\|\mathbf{M}\|_{2, \infty} = \max_i \| \mathbf{M}_{i,\cdot} \|_2$, with $\mathbf{M}_{i,\cdot}$ denoting the $i$-th row of $\mathbf{M}$. For two sequences of nonnegative numbers $\{a_n\}_{n\geq 1}$ and $\{b_n\}_{n\geq 1}$, we write $a_n \lesssim b_n$ to indicate that there exists a constant $C>0$ such that  $a_n \leq Cb_n$ for  sufficiently large $n$.   We denote by $a_n\asymp b_n $ when $a_n\lesssim b_n $ and $b_n\gtrsim a_n$. For an operator $C$, we denote $\overline{\operatorname{Ran}(C)}$ as the closure of the range of $C$. For two Hilbert spaces $\cH_1 $ and \( \cH_2 \),  with inner products $\langle\cdot, \cdot\rangle_{{\cal H}_1}$ and  $\langle\cdot, \cdot\rangle_{{\cal H}_2}$ respectively, we denote their tensor product space as  $\cH_1 \otimes \cH_2$, which is the completion of all finite linear combinations of elementary tensors $\{ \psi  \otimes \phi: \psi \in \cH_1, \phi  \in \cH_2 \}$, with the inner product defined as $\langle \psi_1 \otimes \phi_1, \psi_2 \otimes \phi_2 \rangle = \langle \psi_1, \psi_2 \rangle_{\cH_1} \cdot \langle \phi_1, \phi_2 \rangle_{\cH_2}$. Note that $\cH_1 \otimes \cH_2$ is also a Hilbert space.

\section{Network-linked heterogeneous graphical model}
\label{sec:mod}

Suppose a training sample $\{\bx_i\}_{i=1}^n$ is observed, each with a different probability density function $p_0^{(i)}(\bx)$ on a compact subset $\cX_d \subseteq \bbR^d$.  The conditional independence structure among the variables in $\bX_i=(X_{i1},\ldots, X_{id})^\top$ can be depicted as  $\cG^{(i)} = (\cV, \cE^{(i)})$, where $\cV=[d]$ represents the common node set and $\cE^{(i)} \subseteq \cV \times \cV$ denotes edge set that may vary from node to node. Specifically, for any given node $i\in[n]$,
\begin{align}
\label{eq:cond_ind}
(j,l) \notin {\cE}^{(i)} \ \  \Longleftrightarrow \ \ X_{ij}  \ind X_{il}\ | \ \bX_{i,-\{j,l\}}, 
\end{align}
where $\bX_{i, -\{j,l\}} =\big \{ X_{ik}:k \in {\cal V}\backslash \{j,l\} \big \}$ consists of all the random variables in $\bX_i$ except for $X_{ij}$ and $X_{il}$.  

Suppose further that $\{\bx_i\}_{i=1}^n$ are linked via an undirected network with an adjacency matrix $\bA =\big (A_{i j} \big )_{i, j=1}^n \in\{0,1\}^{n \times n}$, where $\bP =\big (p_{i j} \big )_{i, j=1}^n \in (0,1)^{n \times n}$ with $p_{i j}=P\left(A_{i j}=1\right)$. To leverage the network structure, we introduce a latent network embedding space, where each node $i$ is embedded with a latent vector $\bfbeta_i \in \cX_m \subseteq \bbR^m$, with $\cX_m $ being a compact subset in $\bbR^m$. We consider an integrated model,
\begin{equation}\label{eqn:int_model}
\sigma( p_{i j})= g(\bfbeta_i, \bfbeta_j) \;\text { and }\; p_0^{(i)}(\bx)=p_0\left(\bx ; \bfbeta_i\right), 
\end{equation}
for any $i, j\in [n]$, where $\sigma(\cdot)$ is a link function, and $g: \cX_m\times \cX_m \rightarrow \bbR$. 

This model renders both the latent structure of $p_{ij}$ and the smoothness of $p_0^{(i)}(\bx)$ via the latent network embedding vectors $\bB=(\bfbeta_1, \dots,\bfbeta_n)^{\top}$. The structure information about each node is essentially captured in the $\bfbeta_i$'s, and the smoothness of $p_0^{(i)}(\bx)$ over the discrete network can be transformed into the well-defined smoothness over the continuous $\bfbeta_i$'s.  Furthermore, many network models can be adopted in \eqref{eqn:int_model}, including the latent space model \citep{hoff2002latent} with $\sigma(x) = \operatorname{logit}(x)$ and $g(\bfbeta_i, \bfbeta_j) = \|\bfbeta_i - \bfbeta_j\|_2^2$, the network embedding model \citep{zhang2022directed} with $\sigma(x) = \operatorname{logit}(x)$ and $g(\bfbeta_i, \bfbeta_j) = \bfbeta_i^\top \bfbeta_j$,  the random dot product graph model \citep{young2007random} with $\sigma(x) = x$ and $g(\bfbeta_i, \bfbeta_j) = \bfbeta_i^\top \bfbeta_j$, and the kernel-based latent space model \citep{tang2013universally} with $\sigma(x) = x$ and $g (\bfbeta_i, \bfbeta_j) = k(\bfbeta_i, \bfbeta_j)$ for some kernel function $k$. 

Note that the conditional independence in \eqref{eq:cond_ind} can be fully captured by the second-order partial derivative of $\log p_0(\bx;\bfbeta)$ with respect to $\bx$ \citep{Dawid1979}. 
More specifically, $(j,l) \notin \cE^{(i)}$ if and only if $\partial_j\partial_l \log p_0(\bx;\bfbeta_i) = 0$ almost surely, where $\partial_j$ denotes the partial derivative with respect to the $j$-th coordinate of $\bx$. Therefore, to estimate the heterogeneous graph structure $\cG^{(i)}$ for each node $i$, it suffices to recover the zeros patterns of $\bOmega^{(i)}=  \{\bOmega^{(i)}_{jl}  \}_{j,l=1}^d \in \bbR^{d \times d}$ with 
\begin{equation}
\label{eq_Omega}
\bOmega^{(i)}_{jl} =  \int_{\cX_d} \Big(\partial_j\partial_l \log p_0(\bx ;\bfbeta_i)\Big)^2 p_0(\bx) d \bx,
\end{equation}
where $p_0(\bx) = \frac{1}{n}\sum_{k=1}^n p_0(\bx;\bfbeta_k)$ is the integrated  marginal density of $\bx$. Denote the Stein's score function of $p_0(\bx; \bfbeta)$ as $\bs_{0}(\bx;\bfbeta)=\nabla_{\bx} \log p_0(\bx;\bfbeta) = \big (\partial_1 \log p_0(\bx;\bfbeta), \ldots, \partial_d \log p_0(\bx;\bfbeta) \big )^\top$ and thus $\partial_j\partial_l \log p_0(\bx;\bfbeta)=\partial_j \{\be_l^\top\bs_{0}(\bx;\bfbeta)\}$, where $\be_l$ denotes the $l$-th natural basis of $\mathbb{R}^d$. It then suffices to estimate $\bs_{0}(\bx;\bfbeta)$ and its partial derivatives.

\section{Estimation}\label{sec_estimation}

Estimation of $\cG^{(i)}$ involves three steps, first the estimated network embedding vectors $\widehat \bB = (\widehat \bfbeta_1, \dots, \widehat \bfbeta_n)^\top$, then the estimated score function $\widehat \bs_{\lambda}(\bx;\bfbeta)$ in a vector-valued RKHS, and finally the estimated heterogeneity edge sets $\widehat \cE^{(i)}$. 

\subsection{Vector-valued RKHS}

 We consider a specified vector-valued RKHS $\cH_K$ induced by the following kernel function,
\begin{equation}\label{eqn:kernel_form}
K\big((\mathbf{x}, \bfbeta),(\mathbf{x}^{\prime}, \bfbeta^{\prime})\big)=k_1(\mathbf{x}, \mathbf{x}^{\prime}) k_2(\bfbeta, \bfbeta^{\prime}) \mathbf{I}_d,
\end{equation}
where $k_1: \mathcal{X}_d \times \mathcal{X}_d \rightarrow \mathbb{R}$ and $k_2:\cX_m \times\cX_m \rightarrow \mathbb{R}$ denote two universal scalar kernels, and $\mathbf{I}_d$ is a $d$-dimensional identity matrix. Clearly, $\cH_K$ consists of functions mapping from $\mathcal{X}_d\times \mathcal{X}_m \rightarrow \mathbb{R}^d$ and can be decomposed into a tensor product of two separated RKHSs for $\mathbf{x}$ and $\bfbeta$, and thus allows us to separate their effects in estimating the node-wise heterogeneous graphical model. 

Lemma \ref{lem:H_k} lays out some properties of this specific $\mathcal{H}_K$. Other details about the general vector-valued RKHS are deferred to the Supplementary Material.

\begin{lemma}\label{lem:H_k}
Let $\cH_K$ denote the RKHS induced by the kernel function $K$ in (\ref{eqn:kernel_form}), then it satisfies that
\begin{itemize}
    \item[(a)] $\cH_K=(\cH_1\otimes \cH_2)\times(\cH_1\otimes \cH_2)\times \dots \times (\cH_1\otimes \cH_2)$, where $\cH_1$ and $\cH_2$ are the RKHSs induced by $k_1$ and $k_2$, respectively, $\times$ denotes the Cartesian product, and $\otimes$ denotes the tensor product.
    
    \item[(b)] $\cH_K$ is dense in the set consisting of all continuous functions mapping from $\mathcal{X}_d\times \cX_m$ to $\mathbb{R}^d$.
\end{itemize}
\end{lemma}

Lemma \ref{lem:H_k}(a) characterizes the structure of $\cH_K$ as a Cartesian product of two separate RKHSs, one for latent vector $\bfbeta$ and the other for $\bx$. This structure implies that for any fixed $\bfbeta \in \cX_m$, a function $\bs \in \cH_K$ is mapped to an element $\bs(\cdot; \bfbeta) \in \cH_1^d$, where $\cH_1^d$ is an RKHS induced by the kernel $k_1(\bx, \bx^{\prime})\bI_d$. 
A detailed verification of this claim is provided in Lemma S5.
As a result, the distance between any two functions $\bs_1, \bs_2 \in \cH_K$, evaluated at any specific $\bfbeta_1$ and $\bfbeta_2$, can be measured by $\|\bs_1(\cdot;\bfbeta_1) - \bs_2(\cdot;\bfbeta_2)\|_{\cH_1^d}$, where $\|\cdot\|_{\cH_1^d}$ is the associated RKHS norm in $\cH_1^d$.
Lemma \ref{lem:H_k}(b) further establishes the universality of $\cH_K$, asserting that any continuous function from $\mathcal{X}_d \times \cX_m$ to $\bbR^d$ can be arbitrarily well approximated by $\cH_K$.

\subsection{Efficient algorithm} 
We now develop an efficient algorithm for estimating the heterogeneous $\cG^{(i)}$'s in the vector-valued RKHS $\cH_K$. The first step of $\widehat \bB$ can be obtained by employing any consistent network estimation method, such as \cite{hoff2002latent}, \cite{ma2020universal}, and \cite{zhang2022directed}. Next, $\bs_{0}(\bx;\bfbeta)$ can be estimated by minimizing the integrated Fisher divergence that 
\begin{align}
\label{eqn:score_1}
\argmin_{\bs \in \cal \cH_K} \frac{1}{n}\sum_{i=1}^n  \bbE_{\bfbeta_i} \big\|\bs(\bX;\bfbeta_i)-\bs_0(\bX;\bfbeta_i)\big\|_2^2,
\end{align}
where $\bbE_{\bfbeta_i}$ denotes the expectation with respect to $\bX$ under $p_0(\bx; \bfbeta_i)$.

\begin{lemma}\label{lem:score}
Suppose that $p_0(\bx; \bfbeta)\rightarrow 0$ as $\|\bx\|_2\rightarrow\infty$ for any $\bfbeta$, \eqref{eqn:score_1} is equivalent to
\begin{align*}
\argmin_{\bs \in \cH_K} J(\bs)=\frac{1}{n} \sum_{i=1}^n  \bbE_{\bfbeta_i} \Big [ \frac{1}{2}\big\|\bs(\bX; \bfbeta_i)\big\|_2^2+ \tr\big(\nabla_{\bx} \bs(\bX;\bfbeta_i)\big) \Big ].
\end{align*}
\end{lemma}

Lemma \ref{lem:score} shows that the minimization task in \eqref{eqn:score_1} enjoys the same property of standard score matching \citep{hyvarinen2005estimation, zhou2020nonparametric}, where the unknown $\bs_0(\bx;\bfbeta)$ can be integrated out in the objective function without compromising its global minimizer. With $J(\bs)$ replaced by its empirical counterpart, $\bs_0(\bx;\bfbeta)$ can be estimated by 
\begin{equation}
\label{obj_fun}
\widehat{\bs}_{\lambda}=\argmin_{\bs \in \cH_K} \ \frac{1}{n} \sum_{i=1}^n \Big (\frac{1}{2} \big\|\bs (\bx_i ; \widehat{\bfbeta}_i ) \big\|_2^2+\tr \big(\nabla_{\bx} \bs (\bx_i ; \widehat{\bfbeta}_i )\big ) \Big )+\frac{\lambda}{2} \|\bs  \|_K^2,
\end{equation}
where $\|\cdot\|_K$ is the associated RKHS norm in $\cH_K$. For any $(\bx,\bfbeta)\in\mathcal{X}_d\times \cX_m$ and $j\in[d]$, let $K_{(\bx, \bfbeta)}^*:\cH_K\rightarrow \bbR^d$ denote the operator satisfying $K_{(\bx, \bfbeta)}^*\bs=\bs(\bx;\bfbeta)$ for any $\bs\in \cH_K$,   let $K_{(\bx, \bfbeta)}$ denote its adjoint operator, and let $\partial_{j} K_{(\bx,\bfbeta)}: \bbR^d\rightarrow \cH_K$ denote  a bounded linear operator that  $(\partial_{j}K_{(\bx,\bfbeta)}\be)(\bx^{\prime},\bfbeta^{\prime})=\partial_{x_j}K((\bx^{\prime},\bfbeta^{\prime}),(\bx,\bfbeta))\be$ for any $(\bx^{\prime},\bfbeta^{\prime})\in\mathcal{X}_d\times \cX_m$ and $\be\in \bbR^d$. The properties of vector-valued RKHS ensure the existence of these operators;  see Section S3 in the Supplementary Material. The optimization task in \eqref{obj_fun} can be substantially simplified by leveraging the representer theorem \citep{kimeldorf1971some}.

\begin{lemma}\label{lem:estimation}
The solution to \eqref{obj_fun} must satisfy that
\begin{align*} 
\widehat{\bs}_{\lambda}=\sum_{i=1}^n\sum_{j=1}^d \widehat{ \bfbeta}_{(i-1)d+j}K_{(\bx_{i}, \hat{\bfbeta}_i)} \be_j -\frac{1}{\lambda n} \partial_jK_{(\bx_{i}, \hat{\bfbeta}_i)} \be_j,
\end{align*}
where $\widehat{\bm{\bfbeta}} = \{\widehat{\bfbeta}_{(i-1)d+j}\}_{i,j=1}^{n,d}\in \bbR^{nd} $ denotes the estimated representer coefficients. Moreover,  $\widehat{\bm{\bfbeta}}$ must satisfy the following linear equation system
\begin{align}\label{eqn:finite}
    \big(\bF+n\lambda {\bI}_{nd}\big){\widehat{\bm{\bfbeta}}}=\frac{\bh}{\lambda},
\end{align}
where $\bF \in \bbR^{nd\times nd}$ with  $F_{(i-1)d+j, \ (i'-1)d+j'} = \be_{j}^{\top} K\big((\bx_{i}, \hat{\bfbeta}_i), (\bx_{i'}, \hat{\bfbeta}_{i'})\big)\be_{j^{\prime}}$ for $i,i'=1,\dots,n$, $j,j'=1, \dots, d$, and $\bh \in\bbR^{nd}$ with $h_{(i-1)d+j} =\frac{1}{n}\sum\limits_{i'=1}^n\sum\limits_{j'=1}^d \partial_{j'} \be_j^\top K\big((\bx_{i}, \hat{\bfbeta}_{i}), (\bx_{i'}, \hat{\bfbeta}_{i'})\big) \be_{j'}$ for $i=1,\dots,n$ and $j=1,\dots,d$.
\end{lemma}
Lemma \ref{lem:estimation} adapts the representer theorem for obtaining  $\widehat{\bs}_\lambda(\bx;\bfbeta)$ in the vector-valued RKHS $\cH_K$, whose representer coefficients are obtained by solving the linear equation system in \eqref{eqn:finite}. This greatly facilitates efficient estimation of $\bs_0(\bx;\bfbeta)$ even for large-scale heterogeneity graphs. When the size of \eqref{eqn:finite} is too large, stochastic gradient descent can also be adopted to alleviate the computational burden further.

Once $\widehat{\bs}_{\lambda}$ is obtained, estimation of $\bOmega^{(i)}_{jl}$ can be directly computed as
\begin{align*} 
\widehat{\bOmega}^{(i)}_{jl}=\frac{1}{n}\sum_{k=1}^n \left (\partial_j\{\be_l^{\top}\widehat{\bs}_{\lambda}(\bx_{k};\widehat{\bfbeta}_i)\} \right)^2,
\end{align*}
and the estimated edge set becomes $\widehat{\cE}^{(i)}= \{(j,l):  \widehat{\bOmega}^{(i)}_{jl}\ge \delta_{n}   \}$
with some pre-specified thresholding value $\delta_n$.

\section{Theoretical guarantees}\label{sec_theo}

This section quantifies the asymptotic behavior of the proposed method in estimating the heterogeneous $\cG^{(i)}$'s, assuring that the true $\cG^{(i)}$'s can be exactly recovered with high probability. 

\begin{asup}\label{asup: B}
    Suppose $\widehat \bB$ is a consistent estimate of $\bB$ in that $\|\widehat{\bB}-\bB\|_{2, \infty} = o_p(1)$.
\end{asup}
Assumption \ref{asup: B} assures that every latent network embedding vector in $\bB$ is well estimated, which is satisfied by many existing network estimation methods, such as \cite{lyzinski2014perfect} and \cite{li2023statistical}.

\begin{asup}\label{asup: bound_on_K}
    Suppose $k_1$ is twice continuously differentiable on $\mathcal{X}_d\times \mathcal{X}_d$, and $k_2$ is fourth-order continuously differentiable on $\mathcal{X}_m\times \mathcal{X}_m$.  Moreover, there exist some positive constants $\kappa_1, \kappa_2, \kappa_3,\kappa_4, \kappa_5,\kappa_6$ such that  $\underset{\bx\in \mathcal{X}_d}{\sup}\ k_1(\bx, \bx)\le \kappa_1$, $ \underset{\bfbeta \in \cX_m}{\sup}k_2(\bfbeta, \bfbeta)\le \kappa_2$, $\underset{\bx, \bx'\in \mathcal{X}_d}{\sup}\partial_{x_j}\partial_{{x}^{'}_j} k_1(\bx, \bx')\le \kappa_3$, for all $j\in[d]$, $\underset{\bfbeta, \bfbeta'\in \cX_m}{\sup}\partial_{\beta_l}\partial_{\beta_l^{\prime}} k_2(\bfbeta, \bfbeta')\le \kappa_4$,   $\underset{\bfbeta, \bfbeta'\in \cX_m}{\sup}\partial_{\beta_r}\partial_{\beta_l}\partial_{\beta_r^{\prime}}\partial_{\beta_l^{\prime}}k_2(\bfbeta, \bfbeta^{\prime})\leq \kappa_5$, and $\underset{\bfbeta, \bfbeta'\in \cX_m}{\sup}\partial_{\beta_l} k_2(\bfbeta, \bfbeta')\le \kappa_6$  for all $l,r\in[m]$.
\end{asup}

The differentiability conditions on $k_1$ and $k_2$ guarantee that all functions in $\cH_K$ are at least twice differentiable with respect to $\bx$ and fourth-order differentiable with respect to $\bfbeta$, which is commonly assumed in the literature \citep{steinwart2008support, Sriperumbudur2017}. A stricter smoothness condition on $k_2$ is necessary to control the effect of estimation error in $\widehat{\bB}$ on the estimation accuracy of  $\widehat{\bs}_{\lambda}$.
Assumption \ref{asup: bound_on_K} also requires that the kernel functions in \eqref{eqn:kernel_form} and their corresponding gradients are bounded, a condition commonly adopted in the kernel methods literature \citep{caponnetto2007optimal}. Assumption \ref{asup: bound_on_K} is satisfied by many commonly used kernels, such as the Gaussian and Laplacian kernels.

\begin{asup}\label{asup: C_B_conv}
     Let $C_{\bB} = \frac{1}{n} \sum_{i=1}^n \bbE_{\bfbeta_i}\big[K_{(\bX, \bfbeta_i)}K^*_{(\bX, \bfbeta_i)}\big]$,  then $C_{\bB}$ converges in operator norm to a compact, self-adjoint, and positive semi-definite operator $C$ as $n$ diverges.  Furthermore, $P_{\bB}(I-P)\xrightarrow{s} 0$, where $P_{\bB}$ and $P$ are projections to $\overline{\operatorname{Ran}(C_{\bB})}$ and $\overline{\operatorname{Ran}(C)}$, respectively, and $\xrightarrow{s}$ denotes the convergence of strong operator topology. 
\end{asup}

The operator $C_{\bB}$ can be regarded as a covariance-type operator in RKHS, whose convergence is then analogous to the widely adopted convergence assumption on the sample covariance matrices in literature \citep{ zhao2006model, zou2006adaptive, li2023statistical}. Strong convergence of the projection $P_{\bB}$ is required to establish $\underset{n\rightarrow \infty}{\lim\sup}\|P_{\bB}\bs_0\|_K\le \|P\bs_0\|_K$, which controls the performance of $\bs_0$ on $\bB$ and is considerably weaker than some commonly used assumptions, such as $\bs_0\in \overline{\operatorname{Ran}(C)}$ \citep{Sriperumbudur2017}. More importantly, Assumption \ref{asup: C_B_conv} is satisfied by many existing network models,  including the stochastic block model (SBM) and the random dot product graph model (RDPG).

\begin{theorem} \label{thm:con_s}
    Suppose Assumptions \ref{asup: B}-\ref{asup: C_B_conv} are met, then there exists a sequence $\lambda_n\rightarrow 0$, such that
    $$
 \sup_{i\in [n]}\big \|\widehat{\bs}_{\lambda_n}(\cdot 
  ; \widehat{\bfbeta}_i)-\bs_0(\cdot   ; \bfbeta_i)\big\|_{\cH_1^d} =o_p(1). 
    $$
\end{theorem}
Theorem \ref{thm:con_s} establishes the uniform consistency of $\widehat{\bs}_{\lambda_n}$ in estimating $\bs_0$ in $\cH_1^d$ across all latent positions. A global convergence rate of $\widehat{\bs}_{\lambda}$ in terms of $\|\widehat{\bs}_{\lambda}-\bs_0\|_K$ can also be established under some additional smoothness assumptions on $\bfbeta$, as illustrated in the theoretical example in Section S2.2 of the Supplementary Material.

\begin{theorem}\label{thm:omg_con}
    Suppose all the assumptions in Theorem \ref{thm:con_s} are met. Then, it holds true that
    \begin{align*}
          \sup_{i\in [n]} \big\|\widehat{\bOmega}^{(i)}-\bOmega^{(i)} \big\|_{\max} = o_p(1).
    \end{align*}
\end{theorem}

Theorem \ref{thm:omg_con} further establishes the strong consistency in estimating $\bOmega^{(i)}$ across all nodes, which plays a crucial role in establishing the exact recovery of the heterogeneous graph structure. The strong consistency is established in terms of the matrix max norm, which is in contrast to the Frobenius norm or the spectral norm in most existing results \citep{zhou2010time, li2020high}. Moreover, Theorem \ref{thm:omg_con} immediately implies that 
    \begin{equation*}
        \bbP \Big (\sup_{i\in [n]} \|\widehat{\bOmega}^{(i)}-\bOmega^{(i)} \|_{\max} \le b_n \Big )\ge 1-a_n,
    \end{equation*}
for some non-negative diminishing sequences $\{a_n\}_{n\geq 1}$ and $\{b_n\}_{n\geq 1}$, which may vary with the smoothness and convergence properties in Assumption \ref{asup: C_B_conv}. A detailed verification is provided in Lemma S10.

\begin{theorem}\label{thm:edge_con}
    Suppose all the assumptions in Theorem \ref{thm:con_s} are met, and  $\underset{(j, l)\in \cE^{(i)}}{\inf}\bOmega^{(i)}_{jl}>2 b_n  $  for all $i \in [n]$. With $\delta_n=b_n$, it holds true that 
    \begin{align*}
        \bbP \Big (\widehat{\cE}^{(i)} = \cE^{(i)} \text{ for all } i\in [n] \Big ) \ge 1-2a_n.
    \end{align*}
\end{theorem}

Theorem \ref{thm:edge_con} ensures that the heterogeneous graph structure can be exactly recovered uniformly over all nodes. It is also important to remark that Theorem \ref{thm:edge_con} holds true without requiring any explicit distributional assumptions, which is in sharp contrast to most existing distribution-based methods \citep{li2020high}. In fact, non-asymptotic versions of Theorems \ref{thm:con_s}--\ref{thm:edge_con} can also be established under some relaxed conditions. This point is further illustrated through two examples, the SBM model in Section \ref{sec:th_example} and the RDPG model in Section S2.2 of the Supplementary Material.

\subsection{A theoretical example}\label{sec:th_example}

We now present a theoretical example to illustrate the established results in Theorems 1-3. Suppose $\{\bx_i\}_{i=1}^n$ are observed with a network structure $\bA$ that follows an SBM model with $m$ communities,  parameterized as $\bbE[\bA]=\bP = \bZ \bD \bZ^\top$, where $\bZ \in \{0,1\}^{n\times m}$ is a membership matrix satisfying $\bZ {\bf 1}_m = {\bf 1}_n$ and $\bD\in \bbR^{m\times m}$ is symmetric and nonsingular. Here, ${\bf 1}_n\in\bbR^n$ is a vector with all elements equal to $1$. Assume that the $m$ communities are asymptotically balanced, and thus the proportion of nodes in the $j$-th community converges to a constant $\pi_j^*$ for any $j \in [m]$, with $\sum_{j=1}^m \pi_j^* = 1$. Define $\Delta = \operatorname{diag}(\sqrt{\pi_1^*}, \dots, \sqrt{\pi_m^*})$, and assume $\Delta\bD\Delta$ is positive-definite with distinct eigenvalues. Let $\bB = (\bfbeta_1 , \dots, \bfbeta_n )^\top \in \bbR^{n \times m}$ satisfy $\bB\bB^\top = \bP$, $\bB^\top \bB$ is diagonal, and the elements of $\bfbeta_1$ are all positive. Furthermore, assume each $\bx_i$ are generated from a probability density $p_0(\bx; \bfbeta_i)$, whose heterogeneous graph structure is determined by the zero pattern of the corresponding $\bOmega^{(i)}$ as defined in \eqref{eq_Omega}.

We estimate $\bB$ using the adjacency spectral embedding  \citep{athreya2018statistical}, which conducts a singular value decomposition of $\bA$ and keeps the first $m$ scaled leading singular eigenvectors. The resulting estimator $\widehat{\bB}$ satisfies $\|\widehat{\bB}-\bB \|_{2,\infty} = o_p (1)$, as detailed in Section S2.1 of the Supplementary Material, thereby verifying Assumption \ref{asup: B}.

Next, we estimate the score function $\bs_0=\nabla_{\bx}\log p_0$  in a vector-valued  RKHS induced by $K=k_1k_2\mathbf{I}_{d}$,  as defined in \eqref{eqn:kernel_form}, where $k_1$ and $k_2$ are Gaussian kernels. This ensures Assumption \ref{asup: bound_on_K} is satisfied, while verifying Assumption \ref{asup: C_B_conv} requires further analysis. Specifically, there exists an orthogonal matrix $\bQ_n$ such that $\bB=\bZ \bD^{1/2} \bQ_n$, leading to $\bB^\top\bB=  \bQ_n^\top \bD^{1/2} \bZ^{\top}\bZ \bD^{1/2} \bQ_n$, which is diagonal and denoted as $\bm \Lambda_n$. By the fact that
$n^{-1}\bQ_n {\bm \Lambda}_n \bQ_n^\top= n^{-1} \bD^{1/2} \bZ^{\top}\bZ\bD^{1/2}\rightarrow \bD^{1/2} \Delta^2 \bD^{1/2}$ and Lemma S9, $\bQ_n$ converges to an orthogonal matrix $\bQ$ composed of the eigenvectors of $\bD^{1/2} \Delta^2 \bD^{1/2}$. Let $\bfbeta_j^*$ denote the $j$-th row of $\bD^{1/2}\bQ$ and define $C = \sum_{j=1}^m \pi_j^* \bbE_{\bfbeta_j^*}[K_{(\bX, \bfbeta_j^*)}K^*_{(\bX, \bfbeta_j^*)}]$. 
It follows from the fact that $n^{-1/2}\|\bZ \bD^{1/2} \bQ_n-\bZ \bD^{1/2} \bQ \|_F  \leq n^{-1/2}  \|\bZ \bD^{1/2}\|_F  \|\bQ_n-\bQ\|_2 \le  \|\bD^{1/2}\|_2\|\bQ_n-\bQ\|_2\rightarrow 0$ and Lemma S8 that  $C_{\bB}$ converges to $C$. Furthermore, $P_{\bB}(I-P)\xrightarrow{s} 0$ is verified in 
Section S2.1 of the Supplementary Material. Putting together, all assumptions are verified, and it then follows from Theorem \ref{thm:edge_con} that the heterogeneous graph structure in $p_0(\bx; \bfbeta_i)$ can be exactly recovered with probability tending to 1.

It is important to remark that Theorems \ref{thm:con_s}--\ref{thm:edge_con} can be extended to their non-asymptotic counterparts. Let $\Tilde{\bs}_{\bB, 0}$ denote the projection of $\bs_0$ onto $\overline{\operatorname{Ran}(C_{\bB})}$, and assume $ \Tilde{\bs}_{\bB, 0}= C_{\bB} g_{\bB}$ for some $g_{\bB}\in\cH_K$ such that $\|g_{\bB}\|_{K}\leq L$, where $L$ is a constant independent of $\bB$. For sufficiently large $n$ and any constant $\tau_1>0$, by taking $\lambda_n = \max\big\{\|\widehat{\bB}-\bB\|_{2, \infty}, n^{-1/2}  (\log n)^{1/2} \big\}^{1/2}$, with probability at least $1-n^{-\tau_1}$,  we have 
\begin{equation*} 
     \begin{split} 
      \sup_{i\in[n] } \big\|\widehat{\bs}_{\lambda_n}(\cdot; \widehat{\bfbeta}_i)-\bs_0(\cdot; \bfbeta_i)\big\|_{\cH_1^d} 
       \leq  \tau_2\max\big\{\|\widehat{\bB}-\bB\|_{2, \infty}, n^{-1/2}  (\log n)^{1/2} \big\}^{1/2},
      \end{split}
\end{equation*} 
where  $\tau_2= 2\sqrt{m\kappa_6}(1+\|\bs_0\|_K)+(4\max\{1,\sqrt{\tau_1}\}+2L) 
  \max\{c_3,1\}\max\{\kappa_1\kappa_2,1\}\tau_3$, with 
 $\tau_3=\max\{\sqrt{\kappa_2}c_3(1+\|\bs_0\|_K), 4 d\sqrt{\kappa_2} (\sqrt{\kappa_2\kappa_3}+ \|\bs_0\|_K), 4d\sqrt{\kappa_2}\}$
 and $c_3 =\max\{d\sqrt{\kappa_3\kappa_4 m}, 2\kappa_1\sqrt{\kappa_2\kappa_4 m}\}$.
 Set $b_n= 
\tau_4 (\tau_2+1)\lambda_n$, where 
  $\tau_4=\max\{8\kappa_2\kappa_3 \|\bs_0\|_{K}^2, \kappa_3(1+8\sqrt{\kappa_2}\|\bs_0\|_K)\}$,
  and assume that $\underset{(j, l)\in \cE^{(i)}}{\inf}\bOmega^{(i)}_{jl}>2 b_n ,$ for all $i\in[n]$. With $\delta_n=b_n$, we have
  \[
  \bbP\Big(\widehat{\cE}^{(i)} = \cE^{(i)}, \text{ for all } i\in[n]\Big)\ge 1-2(d^2n^{-1}+n^{-\tau_1}).
  \]
  More details and proof are provided in Section S2.1 of the Supplementary Material.

\section{Numerical experiments}\label{sec:num}

This section investigates the numerical evaluation of the proposed network-linked graphical model, denoted as NGM, and compares it against some existing competitors, including the Graphical lasso [Glasso; \citealt{friedman2008sparse}], the Gaussian graphical model with network cohesion [GNC; \citealt{li2020high}], and the Dynamic graphical model [DGM; \citealt{yang2020estimating}]. We implemented NGM in Python, Glasso using the Python package \texttt{scikit-learn}, and DGM with the R package \texttt{loggle}. The code for GNC is available online. For NGM, we set $k_1(\bx, \bx^{\prime})=\exp \{-\sigma^2\|\bx-\bx^{\prime}\|_2^2\}\bI_d$ and $ k_2(\bfbeta, \bfbeta^{\prime})=\exp \{-\sigma^2\|\bfbeta-\bfbeta^{\prime}\|_2^2\}\bI_m$ with $\sigma$ set as the inverse of the median of all pairwise distances, and $\delta_n$ and $\lambda$ are tuned via 5-fold cross validation. For Glasso, NGM, and DGM, all the tuning parameters or thresholding values are selected via 5-fold cross validation. 

The numerical evaluation of all the methods is assessed via multiple evaluation metrics, including the false positive rate (FPR), true positive rate (TPR), F1-score, Matthews correlation coefficient (MCC), and structural Hamming distance (SHD). Specifically, we denote a correct prediction of the presence or absence of an edge as true positive (TP) or true negative (TN), and a false prediction of the presence or absence of an edge as false positive (FP) or false negative (FN), respectively. Then, the evaluation metrics are computed as
$\text{FPR}=\frac{\FP}{\FP+\TN}, \text{TPR}=\frac{\TP}{\TP+\FN},$  $\text{F1-score}=\frac{2\TP}{2\TP+\FP+\FN}$, and
\begin{equation*}
\operatorname{MCC }= \frac{\mathrm{TP} \times \mathrm{TN}-\mathrm{FP} \times \mathrm{FN}}{\sqrt{(\mathrm{TP}+\mathrm{FP})(\mathrm{TP}+\mathrm{FN})(\mathrm{TN}+\mathrm{FP})(\mathrm{TN}+\mathrm{FN})}}.
\end{equation*}
Moreover, SHD counts the number of edge insertions, deletions, or flips to transform one graph to the other, which is also divided by the total number of possible edges to be normalized to within $[0, 1]$. Note that a good estimation is indicated with small values of FPR and SHD, or large values of TPR, F1-score, and MCC.

\subsection{Simulated examples}

Three simulated examples are considered, each with a different graph structure over a different network structure. The key components of the data generating scheme for each example are provided below, and more details are deferred to Section S1 in the Supplementary Material.

{\bf Example 1. (RDPG with Gaussian)} The data generating scheme follows that in \cite{li2020high}, where the graph structure remains unchanged over all nodes. Particularly, $\bX_i=(X_{i1}, \dots, X_{id})^{\top}$ are sampled independently from  $N(\bmu_i, \bSigma)$ for $i \in [n]$, with $\bmu_i$ and $\bSigma$ to be generated. First, $\bmu_i$'s are generated over a random dot product graph $\bA$, with $\bbE[\bA]=\bP$ and $\bP= a\bB\bB^\top$. Here the elements of $\bB $ are drawn from a uniform distribution on $[0, 1]$, and $a$ is a scaling factor ensuring $\| \bP \|_{\max} \le 1$. Let $\mathbf{M}\in \bbR^{n \times d}$, whose columns are sampled randomly from the first five leading eigenvectors of $\bA$, and let $\bmu_i$ be the $i$-th row of $\mathbf{M}$. Next, we generate an Erd\H{o}s-R\'{e}nyi network $\bS$ with $d$ nodes and a homogeneous connection probability $0.01$, and define $\bSigma$ such that $\bSigma^{-1} = b \big (0.3 \bS + (0.3 |\lambda_{\min}(\bS)| +0.1)I_d \big )$, where $\lambda_{\min}(\bS)$ denotes the smallest eigenvalue of $\bS$, and the scalar $b$ is chosen so that all diagonal elements of $\bSigma^{-1}$ equal to $1$.

{\bf Example 2. (Dynamic graph with Butterfly)} The data generating scheme follows a dynamic graphical model \citep{zhou2010time, yang2020estimating} evolving over the time interval $[0, 1]$. Particularly,  $\bX_i=(X_{i1}, \dots, X_{id})^{\top}$ at the time point $t_i=\frac{i-1}{n}$ is sampled from a dynamic butterfly process \citep{baptista2021learning}, where each pair $\big (X_{i, 2j-1}, X_{i, 2j} \big )$  is independently generated from $X_{i,2j-1}\sim N(0, 1)$ and $X_{i,2j}=X_{i,2j-1}*\text{Bernoulli}\big (p_j(t_i) \big )+\big (1-\text{Bernoulli}(p_j(t_i)) \big )*N(0, 1)$. It can be verified that $\big (X_{i, 2j-1}, X_{i,2j} \big )$ are conditionally dependent given all other variables if and only if $p_j(t_i)\ne 0$, whereas all other pairs of variables are conditionally independent. This property leads to a dynamic bipartite graph, where the edge set at $t_i$ is $\big \{(2j-1, 2j): p_j(t_i)\ne 0 \mbox{ for any } j=1,\dots, d/2 \big \}$, and then the dynamic graph structure can be determined by examining the zero pattern of $\bp(t) = \big (p_1(t), \dots, p_{d/2}(t) \big )^{\top}$.  We further divide the time interval $[0, 1]$ into five equal sub-intervals, and generate $\bp(t)$ such that its zero pattern remains unchanged within each sub-interval. At each sub-interval, $60\%$ element of $\bp(t)$ is set to be non-zero, and the adjacent sub-interval shares $40\%$ common elements. Note that this time-dependent sample can be structured as a chain network structure $\bA$, with $\bbE [\bA]=\bP=(p_{ij})$, $p_{ij}={1}_{\{|i-j|=1\}}$, and $\bB=(t_1, \dots, t_n)^\top$.

{\bf Example 3. (Permutation graph with Laplacian)} In this example, we generate $\bX_i = (X_{i1}, \dots, X_{id})^{\top}$ from a chain Laplacian distribution associated with a network of permutations. Each node $i$ in the network corresponds to a permutation $\bm{\sigma}_i = (\sigma_i(1), \dots, \sigma_i(d))^{\top}$. The expected value of adjacency matrix $\bA$, denoted as $\mathbb{E}[\bA] = \bP$, has its $ij$-th entry $p_{ij}$ representing the scaled Kendall's Tau distance between $\bm{\sigma}_i$ and $\bm{\sigma}_j$. For each node $i$, $\bX_i$ is generated as follows: $X_{i,\sigma_i(1)}$ is sampled from a Laplace distribution with parameters $(0, 1)$, and for $j = 2, \dots, d$, $X_{i,\sigma_i(j)}$ is sampled from a Laplace distribution with parameters $(X_{i, \sigma_i(j-1)}, 1)$. This results in a chain-like graph, with an edge set $\{(\sigma_i(j), \sigma_i(j+1)): j = 1, \dots, d-1\}.$ The permutations $\{\bm{\sigma}_i\}_{i=1}^n$ are generated sequentially: $\bm{\sigma}_1$ is uniformly generated from all permutations of $[d]$, and each subsequent $\bm{\sigma}_{i+1}$ is obtained by swapping randomly two adjacent elements of $\bm{\sigma}_i$.

For each example, the averaged evaluation metrics over 50 replications of all the methods are reported in Tables \ref{table_example1_eta0}-\ref{table_example3_eta0}. Note that if a method fails to detect any edges, its performance is denoted as ``$-$"; and if its running time exceeds 24 hours, its performance is denoted as ``$*$".

\begin{table}[!ht] 
\small
\caption{The averaged evaluation metrics over 50 replications of all the methods in Example 1 together with their standard errors in parentheses.}
\label{table_example1_eta0}
\centering
\medskip
\scalebox{0.7}{
\begin{tabular}{cccccccc}
  \toprule
$(n, d)$ & Method & FPR & TPR & F1-score & SHD & MCC  \\
  \midrule
$(1000, 10)$ 
& \text{NGM} &\pmb{0.006(0.003)} & 0.957(0.021) & \pmb{0.921(0.027)} & \pmb{0.008(0.003)} & \pmb{0.931(0.023)} \\
& \text{GNC} & 0.150(0.010) & 0.860(0.045) & 0.277(0.023) & 0.150(0.010) & 0.337(0.026) \\
& \text{Glasso} & 0.338(0.026)& \pmb{1.000(0.000)}&0.172(0.014)&0.328(0.025)&0.249(0.014) \\

& \text{DGM} & 0.156(0.008)&0.980(0.014)&0.283(0.014)&0.152(0.008)&0.367(0.013) \\
\midrule
$(1000, 100)$ 
& \text{NGM} & \pmb{0.000(0.000)} & \pmb{0.991(0.002)} & \pmb{0.995(0.001)} & \pmb{0.000(0.000)} & \pmb{0.995(0.001)}\\
& \text{GNC} & 0.004(0.001)&0.012(0.007)&0.016(0.003)&0.014(0.001)&0.014(0.004) \\
& \text{Glasso} & 0.151(0.002)&0.794(0.013)&0.095(0.002)&0.152(0.002)&0.175(0.004) \\

& \text{DGM} & 0.083(0.000) & 0.775(0.013) & 0.153(0.003) & 0.085(0.000) & 0.238(0.004)
 \\
\midrule 
$(2000, 1000)$ 
& \text{NGM} & \pmb{0.000(0.000)}& \pmb{0.980(0.001)}& \pmb{0.989(0.001)}& \pmb{0.000(0.000)}& \pmb{0.989(0.001)} \\
& \text{GNC} & $-$  & $-$ & $-$  & $-$ & $-$ \\
& \text{Glasso} & 0.051(0.000)&0.076(0.002)&0.025(0.001)&0.060(0.000)&0.011(0.001)  \\
& \text{DGM} &   $*$  & $*$ & $*$  & $*$ & $*$ \\
\bottomrule   
\end{tabular}
}
\end{table}

\begin{table}[!ht] 
\small
\caption{The averaged evaluation metrics over 50 replications of all the methods in Example 2 together with their standard errors in parentheses. }
\label{table_example2_eta0}
\centering
\medskip
\scalebox{0.7}{
\begin{tabular}{cccccccc}
  \toprule
$(n, d)$ & Method & FPR & TPR & F1-score & SHD & MCC  \\
  \midrule
$(1000, 10)$ 
& \text{NGM} & 0.039(0.002) & 0.940(0.017) & 0.759(0.013) & 0.040(0.002) & 0.756(0.014)\\
& \text{GNC} &0.019(0.003)&0.680(0.040)&0.693(0.035)&0.039(0.004)&0.678(0.037) \\
& \text{Glasso} &0.471(0.044)&0.993(0.007)&0.355(0.037)&0.440(0.041)&0.359(0.038) \\
& \text{DGM} & \pmb{0.024(0.003)} & \pmb{0.960(0.015)} & \pmb{0.843(0.013)} & \pmb{0.025(0.002)} & \pmb{0.844(0.013)}
  \\
\midrule

$(1000, 100)$ 
& \text{NGM} & 0.002(0.000) & 0.817(0.009) & 0.768(0.007) & \pmb{0.002(0.000)}&0.768(0.008) \\
& \text{GNC} & \pmb{0.001(0.000)}&0.673(0.025)&0.737(0.022)&0.003(0.000)&0.740(0.022) \\
& \text{Glasso} & 0.099(0.013)&0.995(0.002)&0.206(0.022)&0.099(0.013)&0.311(0.021) \\
& \text{DGM} & 0.002(0.000)& \pmb{0.919(0.009)}& \pmb{0.835(0.007)}&\pmb{0.002(0.000)}&\pmb{0.840(0.006)} \\
\midrule

$(2000, 1000)$ 
& \text{NGM} & \pmb{0.000(0.000)} & \pmb{0.929(0.002)} & \pmb{0.781(0.001)} & \pmb{0.000(0.000)} & \pmb{0.791(0.001)}\\
& \text{GNC} & 0.000(0.000)&0.246(0.007)&0.391(0.010)&0.000(0.000)&0.484(0.010) \\
& \text{Glasso} & 0.012(0.006)&0.995(0.002)&0.298(0.042)&0.012(0.006)&0.399(0.045)  \\
& \text{DGM} &   $*$  & $*$ & $*$  & $*$ & $*$  \\

\bottomrule   
\end{tabular}
}
\end{table}

\begin{table}[!ht] 
\small
\caption{The averaged evaluation metrics over 50 replications of all the methods in Example 3 together with their standard errors in parentheses. }
\label{table_example3_eta0}
\centering
\medskip
\scalebox{0.7}{
\begin{tabular}{cccccccc}
  \toprule
$(n, d)$ & Method & FPR & TPR & F1-score & SHD & MCC  \\
  \midrule
$(1000, 10)$ & \text{NGM}&  0.615(0.049) & 0.806(0.036) & \pmb{0.382(0.006)} & \pmb{0.531(0.032)} & \pmb{0.177(0.011)}\\
& \text{GNC} &0.994(0.004)&\pmb{0.900(0.011)}&0.306(0.004)&0.816(0.004)&-0.250(0.037)\\
& \text{Glasso} & 0.835(0.011)&0.883(0.021)&0.338(0.008)&0.691(0.011)&0.052(0.031)\\
& \text{DGM} & \pmb{0.554(0.017)} & 0.517(0.044) & 0.274(0.021) & 0.540(0.016) & -0.030(0.037) \\
\midrule

$(1000, 100)$ & \text{NGM} & \pmb{0.152(0.006)} & \pmb{0.773(0.007)} & \pmb{0.176(0.005)} & \pmb{0.154(0.006)} & \pmb{0.241(0.005)} \\
& \text{GNC} & 0.403(0.006)&1.000(0.000)&0.092(0.001)&0.395(0.006)&0.170(0.002) \\
& \text{Glasso} &0.161(0.008)&0.420(0.022)&0.086(0.005)&0.168(0.009)&0.096(0.005) \\
& \text{DGM} & 0.500(0.003) & 0.689(0.017) & 0.053(0.001) & 0.496(0.003) & 0.053(0.005) \\
\midrule

$(2000, 1000)$ 
& \text{NGM} & \pmb{0.012(0.000)} & \pmb{0.983(0.000)} & \pmb{0.240(0.002)} & \pmb{0.012(0.000)} & \pmb{0.364(0.002)} \\
& \text{GNC} & $*$  & $*$ & $*$  & $*$ & $*$ \\
& \text{Glasso} &  0.012(0.002)&0.114(0.014)&0.029(0.003)&0.014(0.002)&0.039(0.004)   \\
& \text{DGM} &  $*$  & $*$ & $*$  & $*$ & $*$   \\
\bottomrule   
\end{tabular}
}
\end{table}

It is evident from Tables \ref{table_example1_eta0}-\ref{table_example3_eta0} that NGM consistently outperforms all other competitors in nearly all scenarios, especially for large-scale graphical models where some other competitors often fail to produce reasonable estimates. Even in Example 2 where the data is generated exactly following DGM, NGM’s performance is still comparable to that of DGM. Also, the computational cost of DGM is substantially more expensive, and it takes longer than 24 hours without producing any output in Example 2 with $d=1000$. The performance of DGM becomes significantly deteriorated in Example 3, where a different data generating scheme with a more flexible network structure is employed. Notably, the performance of GNC and Glasso is less competitive in most scenarios, likely due to their heavy dependence on the Gaussian assumption.

\subsection{Statistician coauthorship dataset}

We now apply NGM to analyze the dataset of research papers published in four leading statistical journals from 2003 to 2012, as described in \cite{ji2016coauthorship}. The dataset is publicly available at \href{https://github.com/ZhengTracyKe/MADStat}{https://github.com/ZhengTracyKe/MADStat}. It includes two coauthorship networks and one citation network among 3607 statisticians. Following \cite{li2020high}, we focus on the second coauthorship network and remove statisticians with one published article, resulting in a network with 635 statisticians. For each statistician, the average frequencies of 300 popular statistical terms are calculated for all their published papers, so that each statistician is represented via a 300-dimensional vector. Furthermore, as analyzed in \cite{ji2016coauthorship}, these authors can be classified into three scientific communities as ``Bayesian", ``Biostatistics" and ``High-dimensional statistics". 

\begin{figure}[!htp]
    \centering
    \includegraphics[width=0.32\linewidth]{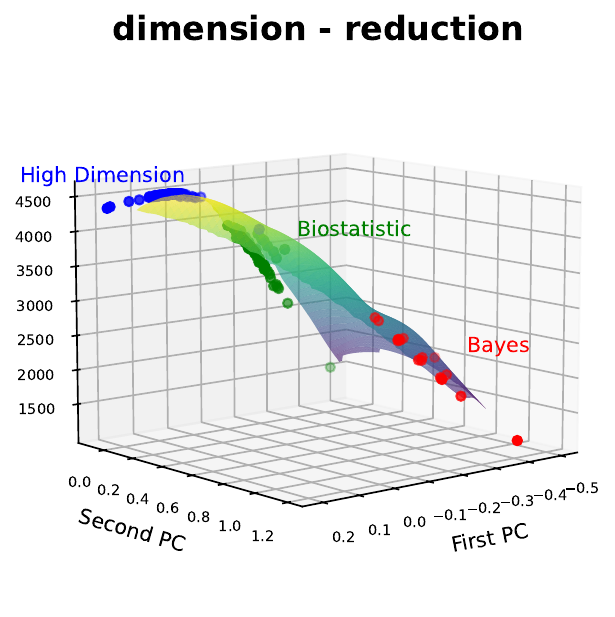}
    \includegraphics[width=0.32\linewidth]{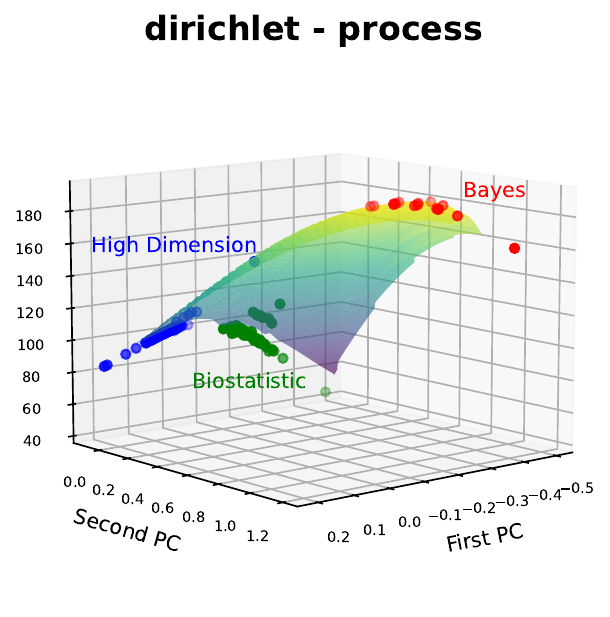}
    \includegraphics[width=0.32\linewidth]{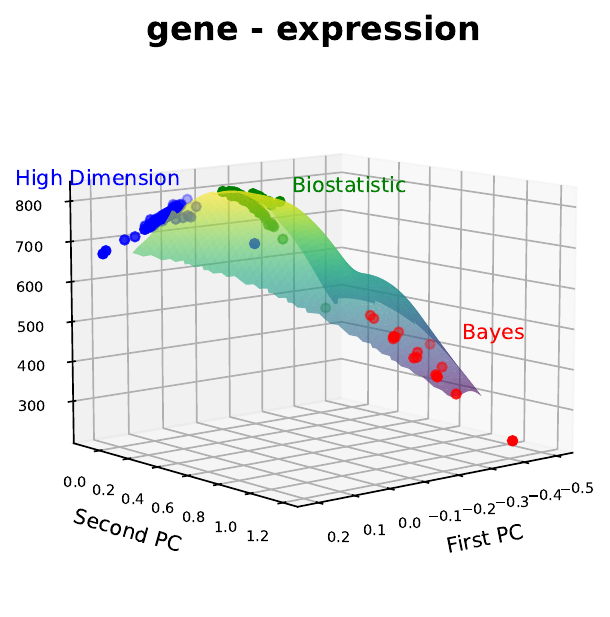}
    \caption{Estimated Signal strengths of three selected statistical term pairs for different statisticians. Each statistician is positioned by their latent embedding vectors and colored by their community membership; green for ``biostatistics", red for ``Bayesian" and blue for ``high dimension". }
    \label{fig:edge_surface}
\end{figure}

We first apply the adjacency spectral embedding method in \cite{athreya2018statistical} to estimate the latent embedding vectors \(\widehat{\bB}\). The proposed NGM is then applied to construct the heterogeneous graphical model based on the strength of $\widehat \bOmega_{jl}^{(i)}$ for each author. Figure \ref{fig:edge_surface} displays the signal strengths of three selected statistical term pairs for the estimated latent embeddings of all statisticians. It is clear that different statistical term pairs exhibit distinct signal strength surfaces. For instance, in the left penal of Figure \ref{fig:edge_surface}, the ``high dimension" community exhibits significantly stronger signal strength of ``dimension - reduction" than those of the other two communities, which is reasonable as dimension reduction has been central to most high-dimensional statistics papers. Similar pattern can be observed in the other two panels, where   ``Bayesian" community exhibits the highest signal strength of ``Dirichlet-process", and ``biostatistics" community exhibits the strongest signal of ``gene-expression". These observations demonstrate that statisticians in different communities possess highly heterogeneous graphical models showing their habits of statistical term usage.

\begin{figure}[!htp]
    \centering
    \includegraphics[width=3.8in]{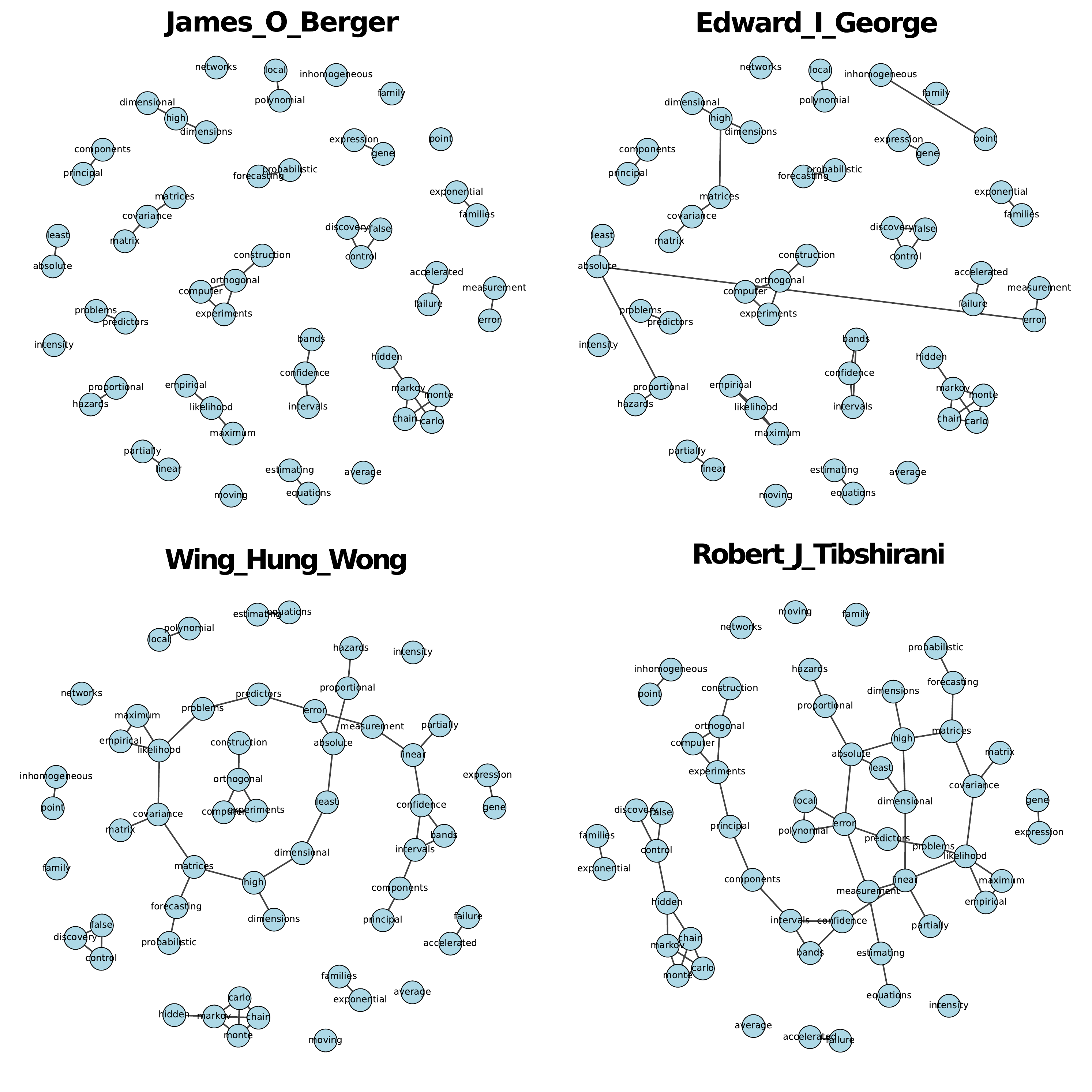}
    \caption{Graphical models of statistical terms for various leading statisticians.}
    \label{fig:graph}
\end{figure}

Figure \ref{fig:graph} displays the estimated graphical models of four leading statisticians on the same 55 terms as selected in \cite{li2020high}. Several common node clusters are present across all four graphical models, such as the cluster consisting of ``Markov", ``chain", ``monte", ``carlo" and ``hidden", and the cluster consisting of ``maximum", ``empirical" and ``likelihood", indicating a strong association among these words across different authors. These common node clusters have also been detected in \cite{li2020high}, which considers a homogeneous graphical model for all statisticians. As a consequence, the homogeneous graphical model fails to detect some personalized patterns for different statisticians, such as ``high - absolute" attains high signal strength for many statisticians in the ``high dimension" community but low signal strength for other statisticians, and thus is not detected in \cite{li2020high}.

More interestingly, it is evident that the four estimated graphical models by NGM in Figure \ref{fig:graph} demonstrate substantial heterogeneity. The graphical models of James O. Berger and Edward I. George exhibit strong similarities, primarily due to their shared membership of the ``Bayesian" community and thus the smoothness of $\widehat \bOmega^{(i)}$ over their latent embedding vectors. The graphical model of Wing Hung Wong substantially differs from those of James O. Berger and Edward I. George and suggests a clear focus on biostatistics research, well aligned with his membership of the ``Biostatistics" community. Finally, though classified into the ``High-dimensional statistics" community, Robert J. Tibshirani's graphical model consists of more edges than other statisticians, due to his diverse research contributions to many statistical communities.

\section{Summary}\label{sec:dis}

This paper proposed a nonparametric framework for learning heterogeneous graphical models on network-linked data, without requiring any specific distributional assumptions. To achieve this, we introduced an efficient two-step estimation algorithm that integrates network embedding with nonparametric graphical model estimation, leveraging vector-valued RKHS for scalable computation. By the representer theorem in RKHS, our nonparametric graphical model estimation leads to a closed-form solution, ensuring both computational efficiency and practical applicability. The proposed method is supported by rigorous theoretical guarantees, ensuring both estimation consistency and exact recovery of the heterogeneous graph structures under mild conditions, which are illustrated through some specific theoretical examples. Numerical experiments on various simulation settings and real data demonstrate that the proposed method outperforms existing competitors in accuracy and scalability. 
 
\if1\blind
{\section*{Acknowledgment}

JW's research is supported in part by HK RGC Grants GRF-11311022, GRF-14306523, GRF-14303424, and CUHK Startup Grant 4937091.
 } \fi

\if0\blind
{ } \fi

\bibliography{ref.bib}

\begin{thebibliography}{}

\bibitem[\protect\citeauthoryear{Athreya, Fishkind, Tang, Priebe, Park, Vogelstein, Levin, Lyzinski, Qin, and Sussman}{Athreya et~al.}{2018}]{athreya2018statistical}
Athreya, A., D.~E. Fishkind, M.~Tang, C.~E. Priebe, Y.~Park, J.~T. Vogelstein, K.~Levin, V.~Lyzinski, Y.~Qin, and D.~L. Sussman (2018).
\newblock Statistical inference on random dot product graphs: a survey.
\newblock {\em Journal of Machine Learning Research,\/}~{\em \textbf{18}\/}(226), 1--92.

\bibitem[\protect\citeauthoryear{Baptista, Morrison, Zahm, and Marzouk}{Baptista et~al.}{2024}]{baptista2021learning}
Baptista, R., R.~Morrison, O.~Zahm, and Y.~Marzouk (2024).
\newblock Learning non-{G}aussian graphical models via {H}essian scores and triangular transport.
\newblock {\em Journal of Machine Learning Research,\/}~{\em \textbf{25}\/}(85), 1--46.

\bibitem[\protect\citeauthoryear{Caponnetto and De~Vito}{Caponnetto and De~Vito}{2007}]{caponnetto2007optimal}
Caponnetto, A. and E.~De~Vito (2007).
\newblock Optimal rates for the regularized least-squares algorithm.
\newblock {\em Foundations of Computational Mathematics,\/}~{\em \textbf{7}\/}(3), 331--368.

\bibitem[\protect\citeauthoryear{Dawid}{Dawid}{1979}]{Dawid1979}
Dawid, A.~P. (1979).
\newblock Conditional independence in statistical theory.
\newblock {\em Journal of the Royal Statistical Society Series B,\/}~{\em \textbf{41}\/}(1), 1--31.

\bibitem[\protect\citeauthoryear{Friedman, Hastie, and Tibshirani}{Friedman et~al.}{2008}]{friedman2008sparse}
Friedman, J., T.~Hastie, and R.~Tibshirani (2008).
\newblock Sparse inverse covariance estimation with the graphical lasso.
\newblock {\em Biostatistics,\/}~{\em \textbf{9}\/}(3), 432--441.

\bibitem[\protect\citeauthoryear{Hoff, Raftery, and Handcock}{Hoff et~al.}{2002}]{hoff2002latent}
Hoff, P.~D., A.~E. Raftery, and M.~S. Handcock (2002).
\newblock Latent space approaches to social network analysis.
\newblock {\em Journal of the American Statistical Association,\/}~{\em \textbf{97}\/}(460), 1090--1098.

\bibitem[\protect\citeauthoryear{Hyv\"{a}rinen}{Hyv\"{a}rinen}{2005}]{hyvarinen2005estimation}
Hyv\"{a}rinen, A. (2005).
\newblock Estimation of non-normalized statistical models by score matching.
\newblock {\em Journal of Machine Learning Research,\/}~{\em \textbf{6}}, 695--709.

\bibitem[\protect\citeauthoryear{Ji and Jin}{Ji and Jin}{2016}]{ji2016coauthorship}
Ji, P. and J.~Jin (2016).
\newblock Coauthorship and citation networks for statisticians.
\newblock {\em The Annals of Applied Statistics,\/}~{\em \textbf{10}\/}(4), 1779--1812.

\bibitem[\protect\citeauthoryear{Kimeldorf and Wahba}{Kimeldorf and Wahba}{1971}]{kimeldorf1971some}
Kimeldorf, G. and G.~Wahba (1971).
\newblock Some results on {T}chebycheffian spline functions.
\newblock {\em Journal of Mathematical Analysis and Applications,\/}~{\em \textbf{33}\/}(1), 82--95.

\bibitem[\protect\citeauthoryear{Kolar and Xing}{Kolar and Xing}{2011}]{Kolar2011}
Kolar, M. and E.~P. Xing (2011).
\newblock On time varying undirected graphs.
\newblock In {\em Proceedings of the Fourteenth International Conference on Artificial Intelligence and Statistics}, pp.\  407--415. PMLR.

\bibitem[\protect\citeauthoryear{Lauritzen}{Lauritzen}{1996}]{Lauritzen1996}
Lauritzen, S.~L. (1996).
\newblock {\em Graphical models}.
\newblock Oxford Statistical Science Series. The Clarendon Press, Oxford University Press, New York.
\newblock Oxford Science Publications.

\bibitem[\protect\citeauthoryear{Le and Li}{Le and Li}{2022}]{le2022linear}
Le, C.~M. and T.~Li (2022).
\newblock Linear regression and its inference on noisy network-linked data.
\newblock {\em Journal of the Royal Statistical Society Series B,\/}~{\em \textbf{84}\/}(5), 1851--1885.

\bibitem[\protect\citeauthoryear{Li, Xu, and Zhu}{Li et~al.}{2023}]{li2023statistical}
Li, J., G.~Xu, and J.~Zhu (2023).
\newblock Statistical inference on latent space models for network data.
\newblock {\em arXiv preprint 2312.06605\/}.

\bibitem[\protect\citeauthoryear{Li, Levina, and Zhu}{Li et~al.}{2019}]{li2019prediction}
Li, T., E.~Levina, and J.~Zhu (2019).
\newblock Prediction models for network-linked data.
\newblock {\em The Annals of Applied Statistics,\/}~{\em \textbf{13}\/}(1), 132--164.

\bibitem[\protect\citeauthoryear{Li, Qian, Levina, and Zhu}{Li et~al.}{2020}]{li2020high}
Li, T., C.~Qian, E.~Levina, and J.~Zhu (2020).
\newblock High-dimensional {G}aussian graphical models on network-linked data.
\newblock {\em Journal of Machine Learning Research,\/}~{\em \textbf{21}\/}(74), 1--45.

\bibitem[\protect\citeauthoryear{Liu, Lafferty, and Wasserman}{Liu et~al.}{2009}]{liu2009nonparanormal}
Liu, H., J.~Lafferty, and L.~Wasserman (2009).
\newblock The nonparanormal: semiparametric estimation of high dimensional undirected graphs.
\newblock {\em Journal of Machine Learning Research,\/}~{\em \textbf{10}}, 2295--2328.

\bibitem[\protect\citeauthoryear{Lu, Kolar, and Liu}{Lu et~al.}{2018}]{lu2018post}
Lu, J., M.~Kolar, and H.~Liu (2018).
\newblock Post-regularization inference for time-varying nonparanormal graphical models.
\newblock {\em Journal of Machine Learning Research,\/}~{\em \textbf{18}\/}(203), 1--78.

\bibitem[\protect\citeauthoryear{Lyzinski, Sussman, Tang, Athreya, and Priebe}{Lyzinski et~al.}{2014}]{lyzinski2014perfect}
Lyzinski, V., D.~L. Sussman, M.~Tang, A.~Athreya, and C.~E. Priebe (2014).
\newblock {Perfect clustering for stochastic blockmodel graphs via adjacency spectral embedding}.
\newblock {\em Electronic Journal of Statistics,\/}~{\em \textbf{8}\/}(2), 2905 -- 2922.

\bibitem[\protect\citeauthoryear{Ma, Ma, and Yuan}{Ma et~al.}{2020}]{ma2020universal}
Ma, Z., Z.~Ma, and H.~Yuan (2020).
\newblock Universal latent space model fitting for large networks with edge covariates.
\newblock {\em Journal of Machine Learning Research,\/}~{\em \textbf{21}\/}(4), 1--67.

\bibitem[\protect\citeauthoryear{Meinshausen and B{\"u}hlmann}{Meinshausen and B{\"u}hlmann}{2006}]{Meinshausen2006}
Meinshausen, N. and P.~B{\"u}hlmann (2006).
\newblock {High-dimensional graphs and variable selection with the Lasso}.
\newblock {\em The Annals of Statistics,\/}~{\em \textbf{34}\/}(3), 1436--1462.

\bibitem[\protect\citeauthoryear{Sriperumbudur, Fukumizu, Gretton, Hyv\"{a}rinen, and Kumar}{Sriperumbudur et~al.}{2017}]{Sriperumbudur2017}
Sriperumbudur, B., K.~Fukumizu, A.~Gretton, A.~Hyv\"{a}rinen, and R.~Kumar (2017).
\newblock Density estimation in infinite dimensional exponential families.
\newblock {\em Journal of Machine Learning Research,\/}~{\em \textbf{18}}, 1--59.

\bibitem[\protect\citeauthoryear{Steinwart and Christmann}{Steinwart and Christmann}{2008}]{steinwart2008support}
Steinwart, I. and A.~Christmann (2008).
\newblock {\em Support Vector Machines}.
\newblock Springer Science \& Business Media.

\bibitem[\protect\citeauthoryear{Tang, Sussman, and Priebe}{Tang et~al.}{2013}]{tang2013universally}
Tang, M., D.~L. Sussman, and C.~E. Priebe (2013).
\newblock Universally consistent vertex classification for latent positions graphs.
\newblock {\em The Annals of Statistics,\/}~{\em \textbf{41}\/}(3), 1406--1430.

\bibitem[\protect\citeauthoryear{Yang and Peng}{Yang and Peng}{2020}]{yang2020estimating}
Yang, J. and J.~Peng (2020).
\newblock Estimating time-varying graphical models.
\newblock {\em Journal of Computational and Graphical Statistics,\/}~{\em \textbf{29}\/}(1), 191--202.

\bibitem[\protect\citeauthoryear{Young and Scheinerman}{Young and Scheinerman}{2007}]{young2007random}
Young, S.~J. and E.~R. Scheinerman (2007).
\newblock Random dot product graph models for social networks.
\newblock In {\em International Workshop on Algorithms and Models for the Web-Graph}, pp.\  138--149. Springer.

\bibitem[\protect\citeauthoryear{Yuan and Lin}{Yuan and Lin}{2007}]{Yuan2007}
Yuan, M. and Y.~Lin (2007).
\newblock {Model selection and estimation in the Gaussian graphical model}.
\newblock {\em Biometrika,\/}~{\em \textbf{94}\/}(1), 19--35.

\bibitem[\protect\citeauthoryear{Zhang, He, and Wang}{Zhang et~al.}{2022}]{zhang2022directed}
Zhang, J., X.~He, and J.~Wang (2022).
\newblock Directed community detection with network embedding.
\newblock {\em Journal of the American Statistical Association,\/}~{\em \textbf{117}\/}(540), 1809--1819.

\bibitem[\protect\citeauthoryear{Zhang, Xu, and Zhu}{Zhang et~al.}{2022}]{zhang2022joint}
Zhang, X., G.~Xu, and J.~Zhu (2022).
\newblock Joint latent space models for network data with high-dimensional node variables.
\newblock {\em Biometrika,\/}~{\em \textbf{109}\/}(3), 707--720.

\bibitem[\protect\citeauthoryear{Zhao and Yu}{Zhao and Yu}{2006}]{zhao2006model}
Zhao, P. and B.~Yu (2006).
\newblock On model selection consistency of {L}asso.
\newblock {\em Journal of Machine Learning Research,\/}~{\em \textbf{7}}, 2541--2563.

\bibitem[\protect\citeauthoryear{Zhou, Lafferty, and Wasserman}{Zhou et~al.}{2010}]{zhou2010time}
Zhou, S., J.~Lafferty, and L.~Wasserman (2010).
\newblock Time varying undirected graphs.
\newblock {\em Machine Learning,\/}~{\em \textbf{80}}, 295--319.

\bibitem[\protect\citeauthoryear{Zhou, Shi, and Zhu}{Zhou et~al.}{2020}]{zhou2020nonparametric}
Zhou, Y., J.~Shi, and J.~Zhu (2020).
\newblock Nonparametric score estimators.
\newblock In {\em International Conference on Machine Learning}, pp.\  11513--11522. PMLR.

\bibitem[\protect\citeauthoryear{Zou}{Zou}{2006}]{zou2006adaptive}
Zou, H. (2006).
\newblock The adaptive lasso and its oracle properties.
\newblock {\em Journal of the American Statistical Association,\/}~{\em \textbf{101}\/}(476), 1418--1429.

\end{thebibliography}
\bibliographystyle{chicago}
\end{document}